\documentclass[aps,preprint]{revtex4}%
\usepackage{amsfonts}
\usepackage{amsmath}
\usepackage{amssymb}
\usepackage{subfigure}
\usepackage{graphicx}%
\begin{document}
\preprint{CTP-SCU/2016002}
\title{Free-fall Frame Black Hole in Gravity's Rainbow}
\author{Jun Tao}
\email{taojun@scu.edu.cn}
\author{Peng Wang}
\email{pengw@scu.edu.cn}
\author{Haitang Yang}
\email{hyanga@scu.edu.cn}
\affiliation{Center for Theoretical Physics, College of Physical Science and Technology,
Sichuan University, Chengdu, 610064, China}

\begin{abstract}
Doubly special relativity (DSR) is an effective model for encoding quantum
gravity in flat spacetime. To incorporate DSR into general relativity, one
could use \textquotedblleft Gravity's rainbow\textquotedblright, where the
spacetime background felt by a test particle would depend on its energy. In
this scenario, one could rewrite the rainbow metric $g_{\mu\nu}\left(
E\right)  $ in terms of some orthonormal frame fields and use the modified
equivalence principle to determine the energy dependence of $g_{\mu\nu}\left(
E\right)  $. Obviously, the form of $g_{\mu\nu}\left(  E\right)  $ depends on
the choice of the orthonormal frame. For a static black hole, there are two
natural orthonormal frames, the static one hovering above it and freely
falling one along geodesics. The cases with the static orthonormal frame have
been extensively studied by many authors. The aim of this paper is to
investigate properties of rainbow black holes in the scenario with the
free-fall orthonormal frame. We first derive the metric of rainbow black holes
and their Hawking temperatures in this free-fall scenario. Then, the
thermodynamics of a rainbow Schwarzschild black hole is studied. Finally, we
use the brick wall model to compute the thermal entropy of a massless scalar
field near the horizon of a Schwarzschild rainbow black hole in this free-fall scenario.

\end{abstract}
\maketitle
\tableofcontents

%\affiliation{Center for Theoretical Physics, College of Physical Science and Technology,
%Sichuan University, Chengdu, 610064, PR China}

%\affiliation{Center for Theoretical Physics, College of Physical Science and Technology,
%Sichuan University, Chengdu, 610064, PR China}

\section{Introduction}

It is generally believed that the framework of the smooth manifold and metric
of classical general relativity breaks down at very high energy scales.
Although a full theory of quantum gravity has yet to available, there are
various attempts using effective models to address this problem. Doubly
Special Relativity (DSR)
\cite{IN-AmelinoCamelia:2000ge,IN-AmelinoCamelia:2000mn,IN-Magueijo:2001cr,IN-Magueijo:2002am}
is one of them, where the non-linear Lorentz transformation in momentum
spacetime is proposed to make the Planck length as a new invariant scale. One
of its predictions is that the transformation laws of special relativity are
modified at very high energies. Thus, the energy-momentum dispersion relation
for a particle of mass $m$ could be modified to%
\begin{equation}
E^{2}f^{2}\left(  E/m_{p}\right)  -p^{2}g^{2}\left(  E/m_{p}\right)  =m^{2},
\label{eq:MDR}%
\end{equation}
where $m_{p}$ is the Planck mass, and $f\left(  x\right)  $ and $g\left(
x\right)  $ are two general functions with the following properties:%
\begin{equation}
\lim_{x\rightarrow0}f\left(  x\right)  =1\text{ and }\lim_{x\rightarrow
0}g\left(  x\right)  =1\text{.}%
\end{equation}
The modified dispersion relation (MDR) might play an important role in
astronomical and cosmological observations, such as the threshold anomalies of
ultra high energy cosmic rays and TeV photons
\cite{IN-AmelinoCamelia:1997gz,IN-Colladay:1998fq,IN-Coleman:1998ti,IN-AmelinoCamelia:2000zs,IN-Jacobson:2001tu,IN-Jacobson:2003bn}%
. One of the popular choice for the functions $f\left(  x\right)  $ and
$g\left(  x\right)  $ has been proposed by Amelino-Camelia et al.
\cite{IN-AmelinoCamelia:1996pj,IN-AmelinoCamelia:2008qg}, which gives%
\begin{equation}
f\left(  x\right)  =1\text{ and }g\left(  x\right)  =\sqrt{1-\eta x^{n}}.
\label{eq:AC-Dispersion}%
\end{equation}
Usually one has $n>0$. As shown in \cite{IN-AmelinoCamelia:2008qg}, this
formula is compatible with some of the results obtained in the
Loop-Quantum-Gravity approach and reflects the results obtained in $\kappa
$-Minkowski and other noncommutative spacetimes. Phenomenological implications
of this \textquotedblleft Amelino-Camelia (AC) dispersion relation" are also
reviewed in \cite{IN-AmelinoCamelia:2008qg}.

To incorporate DSR into the framework of general relativity, Magueijo and
Smolin \cite{IN-Magueijo:2002xx} proposed the \textquotedblleft Gravity's
rainbow", where the spacetime background felt by a test particle would depend
on its energy. Consequently, the energy of the test particle deforms the
background geometry and hence the dispersion relation. As regards the metric,
it would be replaced by a one parameter family of metrics which depends on the
energy of the test particle, forming a \textquotedblleft rainbow\ metric".
Specifically, for the energy-independent metric given by%
\begin{equation}
d\tilde{s}^{2}=\tilde{g}_{\mu\nu}dx^{\mu}\otimes dx^{v}, \label{eq:metric}%
\end{equation}
we could rewrite it in terms of a set of energy-independent orthonormal frame
fields $\tilde{e}_{a}$:%
\begin{equation}
d\tilde{s}^{2}=\eta^{ab}\tilde{e}_{a}\otimes\tilde{e}_{b}.
\end{equation}
Thus, the rainbow modified equivalence principle \cite{IN-Magueijo:2002xx}
implies that the energy-dependent rainbow counterpart for the
energy-independent metric $\left(  \ref{eq:metric}\right)  $ is given by%
\begin{equation}
ds^{2}=\eta^{ab}e_{a}\otimes e_{b},
\end{equation}
where the energy-dependent frame fields are%
\begin{equation}
e_{0}=\frac{\tilde{e}_{0}}{f\left(  E/m_{p}\right)  }\text{ and }e_{i}%
=\frac{\tilde{e}_{i}}{g\left(  E/m_{p}\right)  }\text{.}%
\end{equation}
Note that the MDR $\left(  \ref{eq:MDR}\right)  $ was considered in
\cite{IN-Magueijo:2002xx}. Let see how this works in an example, a static
black hole with the line element%
\begin{equation}
d\tilde{s}^{2}=B\left(  r\right)  dt^{2}-\frac{dr^{2}}{B\left(  r\right)
}-C\left(  r^{2}\right)  h_{\alpha\beta}\left(  x\right)  dx^{\alpha}%
dx^{\beta}, \label{eq:EI-metric}%
\end{equation}
where we assume that the black hole is asymptotically flat which gives
$B\left(  r\right)  \rightarrow1$ as $r\rightarrow\infty$. There are many
choices for $\tilde{e}_{a}$, but one obvious one:%
\begin{equation}
\tilde{e}_{0}=\sqrt{B\left(  r\right)  }dt,\text{ }\tilde{e}_{r}=\frac
{dr}{\sqrt{B\left(  r\right)  }},\text{ and }\tilde{e}_{j},
\label{eq:Static-Tetrad}%
\end{equation}
where $\tilde{e}_{i}$ are some set of one-forms such that $\delta^{ij}%
\tilde{e}_{i}\otimes\tilde{e}_{j}=C\left(  r^{2}\right)  h_{\alpha\beta
}\left(  x\right)  dx^{\alpha}dx^{\beta}$. Therefore, the corresponding
rainbow metric is%
\begin{equation}
ds^{2}=\eta^{ab}e_{a}\otimes e_{b}=\frac{B\left(  r\right)  }{f^{2}\left(
E/m_{p}\right)  }dt^{2}-\frac{dr^{2}}{g^{2}\left(  E/m_{p}\right)  B\left(
r\right)  }-\frac{C\left(  r^{2}\right)  h_{\alpha\beta}\left(  x\right)
dx^{\alpha}dx^{\beta}}{g^{2}\left(  E/m_{p}\right)  }. \label{eq:SF-RB-Metric}%
\end{equation}
For $B\left(  r\right)  =1-\frac{2GM}{r}$ and $C\left(  r^{2}\right)
h_{\alpha\beta}\left(  x\right)  dx^{\alpha}dx^{\beta}=r^{2}d\Omega^{2}$, eqn.
$\left(  \ref{eq:SF-RB-Metric}\right)  $ gives the rainbow Schwarzschild
metric, which was also obtained in \cite{IN-Magueijo:2002xx} using Birkhoff's theorem.

The orthonormal frame adopted in eqn. $\left(  \ref{eq:Static-Tetrad}\right)
$ is a static frame which is anchored to observers hovering above the black
hole. The energy and momentum measured by the static observers would satisfy
the MDR $\left(  \ref{eq:MDR}\right)  $ in the rainbow metric $\left(
\ref{eq:SF-RB-Metric}\right)  $. This rainbow metric $\left(
\ref{eq:SF-RB-Metric}\right)  $ has received a lot of attention and some
relevant work can be found in
\cite{IN-Ling:2005bp,In-Galan:2006by,IN-Li:2008gs,IN-Garattini:2009nq,IN-Salesi:2009kd,IN-Esposito:2010pg,IN-Ali:2014xqa,IN-Gim:2014ira}%
. However, another natural choice for the orthonormal frame is the one
anchored to freely falling observers along the radial direction. For the
energy-independent metric $\left(  \ref{eq:EI-metric}\right)  $, it is obvious
that different choice of orthonormal frame could lead to different form of the
rainbow counterpart. Actually, in section \ref{Sec:FFRBH} we will show that
the rainbow black hole obtained using the free-fall orthonormal frame is given
by%
\begin{equation}
ds^{2}=\frac{dt_{p}^{2}}{f^{2}\left(  E/m_{p}\right)  }-\frac{\left[
dr-v\left(  r\right)  dt_{p}\right]  ^{2}}{g^{2}\left(  E/m_{p}\right)
}-\frac{C\left(  r^{2}\right)  h_{\alpha\beta}\left(  x\right)  dx^{\alpha
}dx^{\beta}}{g^{2}\left(  E/m_{p}\right)  } \label{eq:FF-Rainbow-BH}%
\end{equation}
where $v\left(  r\right)  =-\sqrt{1-B\left(  r\right)  }$ and $t_{p}$ is given
in eqn. $\left(  \ref{eq:tandv}\right)  $. In what follows, we will refer to
the rainbow black holes $\left(  \ref{eq:SF-RB-Metric}\right)  $ and $\left(
\ref{eq:FF-Rainbow-BH}\right)  $ as Static Frame (SF) and Free-fall Frame (FF)
rainbow black holes, respectively.

In this paper, we aim to explore thermodynamics of FF rainbow black holes. For
the static black hole $\left(  \ref{eq:EI-metric}\right)  $, its SF and FF
rainbow counterparts could lead to quite different physics. In the following
sections, we find that

\begin{enumerate}
\item For a test particle, the position of the event horizon of the FF rainbow
black hole $\left(  \ref{eq:FF-Rainbow-BH}\right)  $ is always energy
dependent, which can be obtained by solving eqn. $\left(
\ref{eq:eventhorizon}\right)  $. However, for the SF one $\left(
\ref{eq:SF-RB-Metric}\right)  $, it is obvious that the event horizon radius
$r_{h}$ is energy independent, which is given by $B\left(  r_{h}\right)  =0$.

\item The effective Hawking temperature of the SF rainbow black hole $\left(
\ref{eq:SF-RB-Metric}\right)  $ is \cite{IN-Ali:2014zea}%
\begin{equation}
T_{h}=T_{0}\frac{g\left(  E/m_{p}\right)  }{f\left(  E/m_{p}\right)  },
\label{eq:Th-SF}%
\end{equation}
where $T_{0}$ is the standard Hawking temperature. For the FF one $\left(
\ref{eq:FF-Rainbow-BH}\right)  $, the effective Hawing temperature is given by
eqn. $\left(  \ref{eq:Eff-Temp}\right)  $. In such case, due to the
complicated expression\ for $r_{h}$, the expression for\ $T_{h}$ is usually
more complex than eqn. $\left(  \ref{eq:Th-SF}\right)  $.\ However, for a FF
rainbow Schwarzschild black hole,\ it shows that the effective Hawking
temperature is
\begin{equation}
T_{h}=T_{0}\frac{g^{3}\left(  E/m_{p}\right)  }{f^{3}\left(  E/m_{p}\right)
}.
\end{equation}

\item Thermodynamics of SF and FF rainbow black holes are thus different.
Specifically, for the AC dispersion relation $\left(  \ref{eq:AC-Dispersion}%
\right)  $, we find that the behaviors of SF and FF rainbow Schwarzschild
black holes during the final stage of evaporation process are dramatically
different for $\eta<0$ and $\frac{2}{3}\leq n\leq2$.\ For\ example, a remnant
exists for the FF black hole while it does not for the SF one\ in the case
with $\eta<0$ and $n=\frac{2}{3}$.\ More discussions can be found in section
\ref{Sec:Con}.
\end{enumerate}

The remainder of our paper is organized as follows. In section \ref{Sec:FFRBH}%
, the metric of a FF rainbow black hole is derived, and its Hawking
temperature is obtained using the Hamilton-Jacobi method. The temperature and
entropy of a FF rainbow Schwarzschild black hole are computed in section
\ref{Sec:TRSBH}. In section \ref{Sec:ERSBH}, we calculate the atmosphere
entropy of a massless scalar field near the horizon of a FF rainbow
Schwarzschild black hole using the brick wall model. Section \ref{Sec:Con} is
devoted to our discussion and conclusions. Throughout the paper we take
Geometrized units $c=G=1$, where the Planck constant $\hbar$ is square of the
Planck mass $m_{p}$.

\section{ Free-fall Frame Rainbow Black Hole}

\label{Sec:FFRBH}

The coordinate used in eqn. $\left(  \ref{eq:EI-metric}\right)  $ is the
Schwarzschild-like one, where the line element is diagonal. However, a more
suitable coordinate for describing a specific family of freely falling
observers is the Painleve-Gullstrand (PG) coordinate
\cite{FFRBH-Corley:1996ar,FF}. The PG coordinate anchored to the freely
falling observers along the radial direction takes the form of%
\begin{equation}
d\tilde{s}^{2}=dt_{p}^{2}-\left[  dr-v\left(  r\right)  dt_{p}\right]
^{2}-C\left(  r^{2}\right)  h_{\alpha\beta}\left(  x\right)  dx^{\alpha
}dx^{\beta}, \label{eq:PE-Coprdinate}%
\end{equation}
where $v\left(  r\right)  $ is the velocity of the free fall observers with
respect to the rest observer and $t_{p}$ measures proper time along them. We
assume $v<0,$ $dv/dr>0$ and $v\rightarrow v_{0}\leq0$ as $r\rightarrow\infty$.
Note that $v<0$ means the infalling observers. For simplicity we specialize to
the particular family of observers with $v_{0}=0$ who start at infinity with a
zero initial velocity. Comparing the vector field of the freely falling
observers in PG and Schwarzschild-like coordinates, we find%
\begin{align}
t_{p}  &  =t+\int\frac{\sqrt{1-B\left(  r\right)  }}{B\left(  r\right)
}dr,\nonumber\\
v\left(  r\right)   &  =-\sqrt{1-B\left(  r\right)  }. \label{eq:tandv}%
\end{align}
Requiring $\tilde{e}_{0}=dt_{p}$, we can easily find that the one-forms
$\tilde{e}_{a}$ for the free-fall orthonormal frame are given by%
\begin{equation}
\tilde{e}_{0}=dt_{p},\text{ }\tilde{e}_{r}=dr-v\left(  r\right)  dt_{p},\text{
and }\tilde{e}_{j},
\end{equation}
where $\delta^{ij}\tilde{e}_{i}\otimes\tilde{e}_{j}=C\left(  r^{2}\right)
h_{\alpha\beta}\left(  x\right)  dx^{\alpha}dx^{\beta}$.

In the context of rainbow gravity, the corresponding energy-independent metric
is
\begin{align}
ds^{2}  &  =\frac{\tilde{e}_{0}\otimes\tilde{e}_{0}}{f^{2}\left(
E/m_{p}\right)  }-\frac{\tilde{e}_{r}\otimes\tilde{e}_{r}+\delta^{ij}\tilde
{e}_{i}\otimes\tilde{e}_{j}}{g^{2}\left(  E/m_{p}\right)  }\nonumber\\
&  =\frac{dt_{p}^{2}}{f^{2}\left(  E/m_{p}\right)  }-\frac{\left[  dr-v\left(
r\right)  dt_{p}\right]  ^{2}}{g^{2}\left(  E/m_{p}\right)  }-\frac{C\left(
r^{2}\right)  h_{\alpha\beta}\left(  x\right)  dx^{\alpha}dx^{\beta}}%
{g^{2}\left(  E/m_{p}\right)  }.
\end{align}
The event horizon $r=r_{h}$ will be at which $g^{rr}$ vanishes:%
\begin{equation}
g^{rr}\left(  r_{h}\right)  =v^{2}\left(  r_{h}\right)  f^{2}\left(
E/m_{p}\right)  -g^{2}\left(  E/m_{p}\right)  =0,
\end{equation}
which leads to
\begin{equation}
B\left(  r_{h}\right)  =1-\frac{g^{2}\left(  E/m_{p}\right)  }{f^{2}\left(
E/m_{p}\right)  }. \label{eq:eventhorizon}%
\end{equation}
It is interesting to note that the position of the event horizon depends on
the energy $E$ for FF rainbow black holes while it does not for SF ones.

We now use the Hamilton-Jacobi method to calculate the Hawking temperature of
the FF rainbow black hole $\left(  \ref{eq:FF-Rainbow-BH}\right)  $. After the
Hawking's original derivation, there have been some other methods proposed to
understand the Hawking radiation. Recently, a semiclassical method of modeling
Hawking radiation as a tunneling process has been developed and attracted a
lot of attention. This method was first proposed by Kraus and Wilczek
\cite{FFRBH-Kraus:1994by,FFRBH-Kraus:1994fj}, which is known as the null
geodesic method. Later, the tunneling behaviors of particles were investigated
using the Hamilton-Jacobi method
\cite{FFRBH-Srinivasan:1998ty,FFRBH-Angheben:2005rm,FFRBH-Kerner:2006vu}. In
the Hamilton-Jacobi method, one ignores the self-gravitation of emitted
particles and assumes that their action satisfies the relativistic
Hamilton-Jacobi equation. The tunneling probability for the classically
forbidden trajectory from inside to outside the horizon is obtained by using
the Hamilton-Jacobi equation to calculate the imaginary part of the action for
the tunneling process.

In \cite{RB-SF}, it has been shown that the Hamilton-Jacobi equations for
massless scalars, spin $1/2$ fermions and vector bosons in the rainbow metric
$ds^{2}=\tilde{g}_{\mu\nu}\left(  E\right)  dx^{\mu}dx^{\nu}$ are all given by%
\begin{equation}
\tilde{g}_{\mu\nu}\left(  E\right)  \partial^{\mu}I\partial^{\nu}I=0,
\label{eq:Hamilton-Jacobi}%
\end{equation}
where $I$ is the tunnelling particle's action. From eqn. $\left(
\ref{eq:Hamilton-Jacobi}\right)  ,$ one finds that the Hamilton-Jacobi
equation for a massless particle in the rainbow metric $\left(
\ref{eq:FF-Rainbow-BH}\right)  $ becomes%
\begin{equation}
f^{2}\left(  E/m_{p}\right)  \left[  \partial_{t_{p}}I+v\left(  r\right)
\partial_{r}I\right]  ^{2}=g^{2}\left(  E/m_{p}\right)  \left[  \left(
\partial_{r}I\right)  ^{2}+\frac{h^{\alpha\beta}\left(  x\right)  \left(
\partial_{\alpha}I\right)  \left(  \partial_{\beta}I\right)  }{C\left(
r^{2}\right)  }\right]  . \label{eq:HJ-Rainbow}%
\end{equation}
To solve the Hamilton-Jacobi equation for the action $I$, we can employ the
following ansatz%
\begin{equation}
I=-Et_{p}+W\left(  r\right)  +\Theta\left(  x\right)  ,
\end{equation}
where $E$ is the particle's energy. Plugging the ansatz into eqn. $\left(
\ref{eq:HJ-Rainbow}\right)  $, we have differential equations for $W\left(
r\right)  $ and $\Theta\left(  x\right)  $:%
\begin{gather}
h^{\alpha\beta}\left(  x\right)  \partial_{\alpha}\Theta\left(  x\right)
\partial_{\beta}\Theta\left(  x\right)  =\lambda,\nonumber\\
p_{r}^{\pm}\equiv\partial_{r}W_{\pm}\left(  r\right)  =\frac{-C\left(
r^{2}\right)  v\left(  r\right)  E\pm C\left(  r^{2}\right)  \sqrt{E^{2}%
\frac{g^{2}\left(  E/m_{p}\right)  }{f^{2}\left(  E/m_{p}\right)  }%
+\frac{\lambda}{C\left(  r^{2}\right)  }\left[  v^{2}\left(  r\right)
-\frac{g^{2}\left(  E/m_{p}\right)  }{f^{2}\left(  E/m_{p}\right)  }\right]
\frac{g^{2}\left(  E/m_{p}\right)  }{f^{2}\left(  E/m_{p}\right)  }}}{C\left(
r^{2}\right)  \left[  \frac{g^{2}\left(  E/m_{p}\right)  }{f^{2}\left(
E/m_{p}\right)  }-v^{2}\left(  r\right)  \right]  }, \label{eq:HJ-Lamda&W}%
\end{gather}
where +/$-$ denotes the outgoing/ingoing solutions and $\lambda$ is a
constant. Using the residue theory for the semi circle around $r=r_{h}$, we
get%
\begin{align}
\operatorname{Im}W_{+}\left(  r\right)   &  =\frac{2\pi}{B^{\prime}\left(
r_{h}\right)  }\frac{g\left(  E/m_{p}\right)  }{f\left(  E/m_{p}\right)
}E,\nonumber\\
\operatorname{Im}W_{-}\left(  r\right)   &  =0.
\end{align}
As shown in \cite{SF}, the probability of a particle tunneling from inside to
outside the horizon is%
\begin{equation}
P_{emit}\propto\exp\left[  -\frac{2}{\hbar}\left(  \operatorname{Im}%
W_{+}-\operatorname{Im}W_{-}\right)  \right]  .
\end{equation}
There is a Boltzmann factor in $P_{emit}$ with an effective Hawking
temperature, which is
\begin{equation}
T_{h}=\frac{\hbar B^{\prime}\left(  r_{h}\right)  }{4\pi}\frac{f\left(
E/m_{p}\right)  }{g\left(  E/m_{p}\right)  }, \label{eq:Eff-Temp}%
\end{equation}
where we take $k_{B}=1$.

\section{Thermodynamics of Rainbow Schwarzschild Black Hole}

\label{Sec:TRSBH}

In this section, for simplicity we consider a FF rainbow Schwarzschild black
hole of mass $M$ with $B\left(  r\right)  =1-\frac{2M}{r}$ in eqn. $\left(
\ref{eq:FF-Rainbow-BH}\right)  $. For the FF rainbow Schwarzschild black hole,
eqn. $\left(  \ref{eq:eventhorizon}\right)  $ gives the position of the event
horizon:%
\begin{equation}
r_{h}=2M\frac{f^{2}\left(  E/m_{p}\right)  }{g^{2}\left(  E/m_{p}\right)  }.
\label{eq:horizon-radius}%
\end{equation}
Thus, eqn. $\left(  \ref{eq:Eff-Temp}\right)  $ leads to the effective Hawking
temperature:%
\begin{equation}
T_{h}=T_{0}\frac{g^{3}\left(  E/m_{p}\right)  }{f^{3}\left(  E/m_{p}\right)
}, \label{eq:Eff-Temp-sc}%
\end{equation}
where $T_{0}=$ $\frac{\hbar}{8\pi M}$.

As in \cite{RB-SF}, the Heisenberg uncertainty principle can be used to
estimate the black hole's temperature. The Heisenberg uncertainty principle
gives a relation between the momentum $p$ of an emitted particle and the event
horizon radius $r_{h}$ of the black hole
\cite{TRSBH-Bekenstein:1973ur,TRSBH-Adler:2001vs}:%
\begin{equation}
p/m_{p}\sim\delta p/m_{p}\sim\hbar/m_{p}\delta x\sim m_{p}/r_{h}.
\label{eq:momentum}%
\end{equation}
Assuming that the emitted particle is massless, we find that the modified
dispersion relation $\left(  \ref{eq:MDR}\right)  $ becomes%
\begin{equation}
\frac{E}{m_{p}}\frac{f\left(  E/m_{p}\right)  }{g\left(  E/m_{p}\right)
}=\frac{p}{m_{p}}. \label{eq:MDR-Massless}%
\end{equation}
Substituting eqn. $\left(  \ref{eq:horizon-radius}\right)  $ into eqn.
$\left(  \ref{eq:momentum}\right)  $ and using eqn. $\left(
\ref{eq:MDR-Massless}\right)  $, we have for the energy of the particle:%
\begin{equation}
x\frac{f^{3}\left(  x\right)  }{g^{3}\left(  x\right)  }=y, \label{eq:h}%
\end{equation}
where $x\equiv E/m_{p}$ and $y\equiv\frac{m_{p}}{2M}$. To express the black
hole's temperature in terms of $M$, one can solve eqn. $\left(  \ref{eq:h}%
\right)  $ for $x$ in terms of $y$. In fact, the solution for $x$ can be
expressed as%
\begin{equation}
x=yh\left(  y\right)  , \label{eq:hFunction}%
\end{equation}
where eqn. $\left(  \ref{eq:h}\right)  $ is inverted to obtain the function
$h\left(  y\right)  $ and $\lim\limits_{y\rightarrow0}h\left(  y\right)  =1$.
Substituting eqn. $\left(  \ref{eq:hFunction}\right)  $ into eqn. $\left(
\ref{eq:Eff-Temp-sc}\right)  $ gives the black hole's temperature:%
\begin{equation}
T_{BH}=T_{0}\frac{x}{y}=T_{0}h\left(  \frac{m_{p}}{2M}\right)  .
\label{eq:BH-Temp}%
\end{equation}
The range of the left-hand side (LHS) of eqn. $\left(  \ref{eq:h}\right)  $
determines the ranges of the values\ of $M$. Specifically, the maximum value
of the LHS of eqn. $\left(  \ref{eq:h}\right)  $, which is denoted by $y_{cr}%
$, gives that $M\geq\frac{m_{p}}{2y_{cr}}$. If $y_{cr}$ is finite, it predicts
the existence of the black hole's remnant. For some functions $f\left(
x\right)  $ and $g\left(  x\right)  $, the domain of the LHS of eqn. $\left(
\ref{eq:h}\right)  $ might be $\left[  0,x_{cr}\right]  $/$\left[
0,x_{cr}\right)  $ with $x_{cr}$ being finite. Thus, it gives that the energy
of the particle $E\leq m_{p}x_{cr}$. If the domain is $\left[  0,\infty
\right)  $, we simply set $x_{cr}=\infty$.

For the AC dispersion relation given in eqn. $\left(  \ref{eq:AC-Dispersion}%
\right)  $, eqn. $\left(  \ref{eq:MDR-Massless}\right)  $ becomes%
\begin{equation}
\frac{x}{\left(  1-\eta x^{n}\right)  ^{\frac{3}{2}}}=y. \label{eq:x&y-AC}%
\end{equation}
If $\eta>0,$ one finds that $y_{cr}=0$. However, there is an upper bound
$x_{cr}=\eta^{-1/n}$ on $x$ to make the LHS of eqn. $\left(  \ref{eq:x&y-AC}%
\right)  $ real. If $\eta<0$, $x_{cr}=\infty$ and $y_{cr}=\infty$ for
$0<n<\frac{2}{3}$, and $x_{cr}=\infty$ and $y_{cr}=\left\vert \eta\right\vert
^{-3/2}$ for $n=\frac{2}{3}$. For the case with $\eta<0$ and $n>\frac{2}{3},$
the LHS of eqn. $\left(  \ref{eq:x&y-AC}\right)  $ has a global maximum value
$y_{0}$ at $x_{0}$, where we define%
\begin{align}
x_{0}  &  \equiv\left(  \frac{2-3n}{2}\eta\right)  ^{-\frac{1}{n}},\nonumber\\
y_{0}  &  \equiv\left(  \frac{3n-2}{3n}\right)  ^{\frac{3}{2}}\left(
\frac{2-3n}{2}\eta\right)  ^{-\frac{1}{n}}.
\end{align}
Thus, it would appear that $y\leq y_{0}$ and $x<\infty$ since $x$ can go to
infinity. However, as argued in \cite{RB-SF,TRSBH-Simon:1990ic}, the
"runaways" solution to eqn. $\left(  \ref{eq:x&y-AC}\right)  $, which does not
exist in the limit of $\eta\rightarrow0$, should be discarded. In this case,
we have $x_{cr}=x_{0}$ instead of $x_{cr}=\infty$. We list $x_{cr}$ and
$y_{cr}$ for various choices of $n$ and $\eta$ in TABLE \ref{tab:one}. If
$y\ll1$, one has $x\ll1$, and hence eqn. $\left(  \ref{eq:x&y-AC}\right)  $
becomes%
\begin{equation}
y=x\left(  1+\frac{3\eta x^{n}}{2}+\mathcal{O}\left(  x^{2n}\right)  \right)
, \label{eq:y}%
\end{equation}
which gives%
\begin{equation}
h\left(  y\right)  =1-\frac{3\eta y^{n}}{2}+\mathcal{O}\left(  y^{2n}\right)
. \label{eq:hFunction-Small}%
\end{equation}
Thus for $M\gg m_{p},$ we have from eqn. $\left(  \ref{eq:hFunction-Small}%
\right)  $ that%
\begin{equation}
T_{BH}=\frac{m_{p}^{2}}{8\pi M}\left[  1-\frac{3\eta}{2^{n+1}}\frac{m_{p}^{n}%
}{M^{n}}+\mathcal{O}\left(  \frac{m_{p}^{2n}}{M^{2n}}\right)  \right]  .
\end{equation}

\begin{table}[pbh]
\begin{center}
$%
\begin{tabular}
[c]{|c|c|c|c|c|c|}\hline
& $x_{cr}$ & $y_{cr}$ & $M_{cr}$ & $T_{BH}^{cr}/m_{p}$ & Lines in
figures\\\hline
$\eta=0$ & $\infty$ & $\infty$ & $0$ & $\infty$ & Blue Solid\\\hline
$\eta>0$ & $\eta^{-1/n}$ & $\infty$ & $0$ & $\frac{\eta^{-1/n}}{4\pi}$ & Black
Solid\\\hline
$\eta<0,0<n<\frac{2}{3}$ & $\infty$ & $\infty$ & $0$ & $\infty$ & Black
Dashed\\\hline
$\eta<0,n=\frac{2}{3}$ & $\infty$ & $\left\vert \eta\right\vert ^{-\frac{3}%
{2}}$ & $\frac{m_{p}\left\vert \eta\right\vert ^{\frac{3}{2}}}{2}$ & $\infty$
& Red Dashed\\\hline
$\eta<0,n>\frac{2}{3}$ & $x_{0}$ & $y_{0}$ & $\frac{m_{p}}{2y_{0}}$ &
$\frac{x_{0}}{4\pi}$ & Red Solid\\\hline
\end{tabular}
$
\end{center}
\caption{The values of $x_{cr}$, $y_{cr}$, $M_{cr}$, and $T_{BH}^{cr}/m_{p}$
for a FF rainbow Schwarzschild black hole.}%
\label{tab:one}%
\end{table}

\begin{figure}[tb]
\begin{centering}
\includegraphics[scale=1.2]{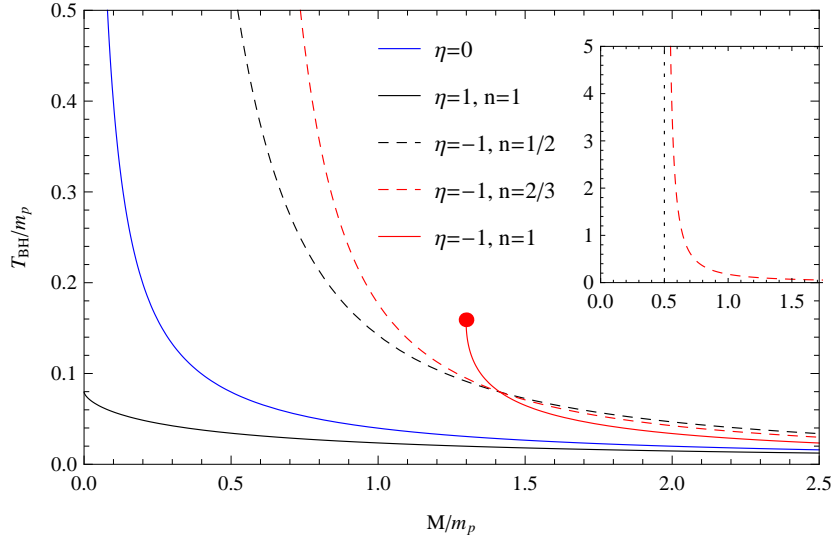}
\par\end{centering}
\caption{Plot of the temperature $T_{BH}/m_{p}$ against the mass $M/m_{p}$ for
a FF rainbow Schwarzschild black hole. All the lines asymptotically approach
$T_{BH}=0$ as $M/m_{p}\rightarrow\infty$. The blue line is the usual case,
where $T_{BH}$ blows up as $M\rightarrow0$. The red dot is the end of the red
solid line, where the black hole has a remnant $M_{cr}=\frac{3^{\frac{3}{2}}%
}{4}m_{p}$. In this case, $T_{BH}$ does not blow up as $M\rightarrow M_{cr}$.
The black dotted line is the asymptotic line of the red dashed line as
$M\rightarrow M_{cr}=0.5m_{p}$, which is the black hole's remnant. In this
case, $T_{BH}$ blows up as $M\rightarrow M_{cr}$.}%
\label{fig:TBH}%
\end{figure}

The minimum mass $M_{cr}$ of the black hole is given by%
\begin{equation}
M_{cr}=\frac{m_{p}}{2y_{cr}}. \label{eq:T-BH}%
\end{equation}
When the mass $M$ reaches $M_{cr},$ the final temperature of the black hole is
denoted by $T_{BH}^{cr}$. Eqn. $\left(  \ref{eq:BH-Temp}\right)  $ gives that%
\begin{equation}
T_{BH}^{cr}=\frac{x_{cr}m_{p}}{4\pi}. \label{eq:Tem-SCBH}%
\end{equation}
For $\eta<0$ and $n\geq\frac{2}{3}$, $y_{cr}$ is finite, and hence the black
hole would have non-vanishing minimum mass $M_{cr}$. This implies the
existence of the black hole's remnant due to rainbow gravity. By eqn. $\left(
\ref{eq:Tem-SCBH}\right)  $, we find that $T_{BH}^{cr}$ is infinite for
$n=\frac{2}{3}$ while $T_{BH}^{cr}$ is $\frac{x_{0}m_{p}}{4\pi}$ for
$n>\frac{2}{3}$. For $\eta<0$ and $0<n<\frac{2}{3}$, we find that $M_{cr}=0$
and $T_{BH}^{cr}=\infty$. In this case, the black hole would evaporate
completely while its temperature increases and finally becomes infinity during
evaporation, just like the standard Hawking radiation. For $\eta>0$, the black
hole would also evaporate completely. However, the temperature of the black
hole is a finite value $\frac{\eta^{-1/n}m_{p}}{4\pi}$ at the end of the
evaporation process. We list $M_{cr}$ and $T_{BH}^{cr}$ for all the possible
values of $\eta$ and $n$ in TABLE \ref{tab:one}. In FIG. $\ref{fig:TBH}$, we
plot the temperature $T_{BH}/m_{p}$ against the black hole mass $M/m_{p}$, for
examples with $\left(  \eta,n\right)  =\left(  1,1\right)  $, $\left(
\eta,n\right)  =\left(  -1,\frac{1}{2}\right)  $, $\left(  \eta,n\right)
=\left(  -1,\frac{2}{3}\right)  $, and $\left(  \eta,n\right)  =\left(
-1,1\right)  $. The standard Hawking radiation is also plotted as a blue line
in FIG. $\ref{fig:TBH}$.

\begin{figure}[tb]
\begin{centering}
\includegraphics[scale=1.2]{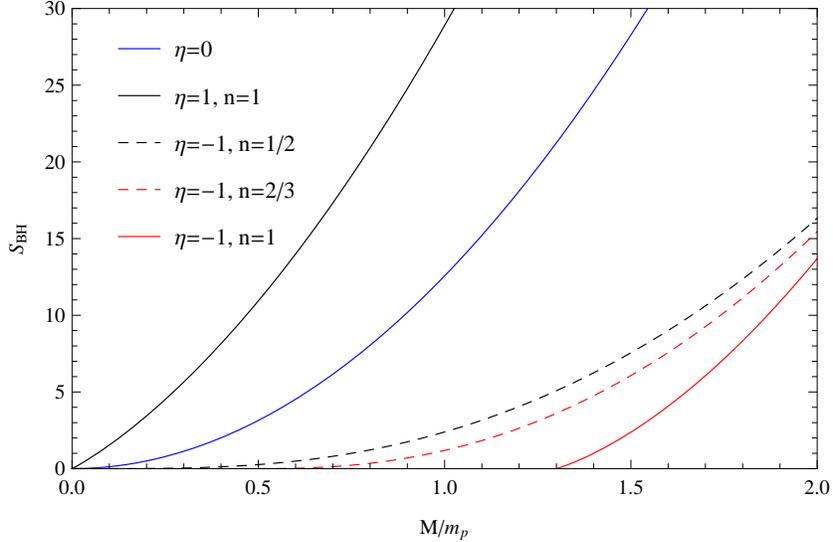}
\par\end{centering}
\caption{Plot of the entropy $S_{BH}$ against the mass $M/m_{p}$ for a FF
rainbow Schwarzschild black hole.}%
\label{fig:SBH}%
\end{figure}

Using the first law of black hole thermodynamics $dS_{BH}=dM/T_{BH}$, we find
that the entropy of the black hole is
\begin{equation}
S_{BH}=\int_{M_{cr}}^{M}\frac{dM}{T_{BH}}=2\pi\int_{\frac{m_{p}}{2M}}^{y_{cr}%
}\frac{dy}{y^{3}h\left(  y\right)  }, \label{eq:Entropy-SC}%
\end{equation}
where $y_{cr}=\frac{m_{p}}{2M_{cr}}$. For the usual case, we have $h\left(
y\right)  =1$ and $y_{cr}=\infty$. Thus, eqn. $\left(  \ref{eq:Entropy-SC}%
\right)  $ gives the Bekenstein-Hawking entropy%
\begin{equation}
S_{BH}=\frac{4\pi M^{2}}{m_{p}^{2}}=\frac{A}{4\hbar}.
\end{equation}
where $A=4\pi\left(  2M\right)  ^{2}$ is the horizon area of the usual
Schwarzschild black hole. If $M\gg m_{p}$ $\left(  A\gg\hbar\right)  $, eqn.
$\left(  \ref{eq:Entropy-SC}\right)  $ gives the entropy up to the subleading
term
\begin{equation}
S_{BH}\sim\left\{
\begin{array}
[c]{l}%
\frac{A}{4\hbar}+\frac{3\pi\eta}{2-n}\left(  \frac{A}{4\pi\hbar}\right)
^{\frac{2-n}{2}}\text{ \ \ \ \ }n\neq2\\
\text{ \ }\frac{A}{4\hbar}+\frac{3\pi\eta}{2}\ln\frac{A}{4\pi\hbar}\text{
\ \ \ \ \ \ \ }n=2
\end{array}
\right.  , \label{eq:Entropy-SC-Small}%
\end{equation}
where we use eqn. $\left(  \ref{eq:hFunction-Small}\right)  $ for $h\left(
y\right)  $. The leading terms of eqn. $\left(  \ref{eq:Entropy-SC-Small}%
\right)  $ are the familiar Bekenstein-Hawking entropy. For $n=2$, we obtain
the logarithmic subleading term. In FIG. $\ref{fig:SBH}$, we plot the entropy
$S$ against the black hole mass $M/m_{p}$, for examples with $\eta=0$,
$\left(  \eta,n\right)  =\left(  1,1\right)  $, $\left(  \eta,n\right)
=\left(  -1,\frac{1}{2}\right)  $, $\left(  \eta,n\right)  =\left(
-1,\frac{2}{3}\right)  $, and $\left(  \eta,n\right)  =\left(  -1,1\right)  $.

\section{Entropy of Rainbow Schwarzschild Black Hole in Brick Wall Model}

\label{Sec:ERSBH}

Although all the evidences suggest that the Bekenstein-Hawking entropy is the
thermodynamic entropy, the statistical origin of black holes' entropy has not
yet been fully understood. One of candidate for the statistical origin is the
entropy of the thermal atmosphere of black holes. However, when one attempts
to calculate the entropy of the thermal atmosphere, there are two kinds of
potential divergences. The first one arises from infinite volume of the
system, which has to do with the contribution from the vacuum surrounding the
system at large distances and is of little relevance here. The second one is
due to the infinite volume of the deep throat region near the horizon. To
regulate the divergences, t' Hooft \cite{ERSBH-'tHooft:1984re} proposed the
brick wall model for a scalar field $\phi$, where two brick wall cutoffs are
introduced at some small distance $r_{\varepsilon}$ from the horizon and at a
large distance $L\gg r_{h}$,
\begin{equation}
\phi=0\text{ \ at }r=r_{h}+r_{\varepsilon}\text{ and }r=L\text{.}
\label{eq:DirichletBoundary}%
\end{equation}
In this section, we will use the brick wall model to calculate the entropy of
a scalar field for a FF rainbow Schwarzschild black hole with $B\left(
r\right)  =1-\frac{2M}{r}$ in eqn. $\left(  \ref{eq:FF-Rainbow-BH}\right)  .$

For particles emitted in a wave mode with energy $E$, one has that%
\begin{align*}
&  \left(  \text{Probability for a black hole to emit a particle in this
mode}\right) \\
&  =\exp\left(  -\frac{E}{T_{h}}\right)  \times(\text{Probability for a black
hole to absorb a particle in the same mode}),
\end{align*}
where $T_{h}$ is given by eqn. $\left(  \ref{eq:Eff-Temp-sc}\right)  $. The
above relation was first obtained by Hartle and Hawking
\cite{ERSBH-Hartle:1976tp} using semiclassical analysis. Neglecting
back-reaction, detailed balance condition requires that the ratio of the
probability of having $N$ particles in a particular mode to the probability of
having $N-1$ particles in the same mode is $\exp\left(  -\frac{E}{T_{h}%
}\right)  .$ The argument in \cite{SF} gives the von Neumann entropy $s_{E}$
for the mode%
\begin{equation}
s_{E}=s\left(  \frac{E}{T_{h}}\right)  , \label{eq:Number}%
\end{equation}
where we define%
\begin{equation}
s\left(  x\right)  =\frac{\left(  -1\right)  ^{\epsilon}\exp x}{\exp x-\left(
-1\right)  ^{\epsilon}}\ln\left[  \frac{\exp x}{\exp x-\left(  -1\right)
^{\epsilon}}\right]  +\frac{\ln\left[  \exp x-\left(  -1\right)  ^{\epsilon
}\right]  }{\exp x-\left(  -1\right)  ^{\epsilon}}.
\end{equation}
Note that $\epsilon=0$ for bosons and $\epsilon=1$ for fermions. As discussed
in section \ref{Sec:TRSBH}, it is interesting to note that there is an upper
bound $m_{p}x_{cr}$ on the energy $E$ of the particle.

For a Schwarzschild black hole, a wave mode of emitted scalars can be labelled
by the energy $E$, angular momentum $l$, and magnetic quantum number $m$.
Thus, the atmosphere entropy of a massless scalar field can be expressed in
the form of%
\begin{equation}
S_{rad}=\int\left(  2l+1\right)  dl\int_{0}^{E_{\max}}dE\frac{dn\left(
E,l\right)  }{dE}s_{E}, \label{eq:entropy-rad}%
\end{equation}
where $E_{\max}=m_{p}x_{cr}$, and $n\left(  E,l\right)  $ is the number of
one-particle states not exceeding $E$ with fixed value of angular momentum
$l$. To obtain $n\left(  E,l\right)  $, we can define the radial wave number
$k\left(  r,l,E\right)  $ by%
\begin{equation}
k^{\pm}\left(  r,l,\omega\right)  =p_{r}^{\pm}, \label{eq:k(r)}%
\end{equation}
as long as $p_{r}^{\pm2}\geq0$, and $k^{\pm}\left(  r,l,E\right)  =0$
otherwise. Note that $p_{r}^{\pm}$ are given in eqn. $\left(
\ref{eq:HJ-Lamda&W}\right)  $, and $\lambda=\left(  l+\frac{1}{2}\right)
^{2}\hbar^{2}$ there for the Schwarzschild black hole\cite{SF}. With these two
Dirichlet boundaries, one finds\cite{FF} that $n\left(  E,l\right)  $ is%
\begin{equation}
n\left(  E,l\right)  =\frac{1}{2\pi\hbar}\left[  \int_{r_{h}+r_{\varepsilon}%
}^{L}k^{+}\left(  r,l,E\right)  dr+\int_{L}^{r_{h}+r_{\varepsilon}}%
k^{-}\left(  r,l,E\right)  dr\right]  . \label{eq:number}%
\end{equation}
Defining%
\begin{equation}
u\equiv\frac{E}{T_{h}}=\frac{E}{T_{0}}\frac{f^{3}\left(  E/m_{p}\right)
}{g^{3}\left(  E/m_{p}\right)  },
\end{equation}
we can use eqns. $\left(  \ref{eq:h}\right)  $ and $\left(  \ref{eq:hFunction}%
\right)  $ to show that%
\begin{equation}
\frac{g\left(  E/m_{p}\right)  }{f\left(  E/m_{p}\right)  }=h^{\frac{1}{3}%
}\left(  \frac{uT_{0}}{m_{p}}\right)  . \label{eq:goverf}%
\end{equation}
Thus, eqn. $\left(  \ref{eq:entropy-rad}\right)  $ becomes%
\begin{align}
S_{rad}  &  =\frac{1}{\hbar^{2}}\int_{0}^{u_{\max}}dus\left(  u\right)
\frac{d}{du}\left[  \int d\lambda n\left(  u,\lambda\right)  \right]
\nonumber\\
&  =\frac{2T_{0}^{3}}{3\pi\hbar^{3}}\int_{0}^{u_{\max}}dus\left(  u\right)
\frac{d}{du}\left[  \int_{r_{h}+r_{\varepsilon}}^{L}dr\frac{r^{2}u^{3}%
h^{\frac{10}{3}}\left(  \frac{uT_{0}}{m_{p}}\right)  }{\left[  B\left(
r\right)  +h^{\frac{2}{3}}\left(  \frac{uT_{0}}{m_{p}}\right)  -1\right]
^{2}}\right]  , \label{eq:entropy-rad1}%
\end{align}
where $u_{\max}=\frac{m_{p}y_{cr}}{T_{0}}$ and $\lambda=\left(  l+\frac{1}%
{2}\right)  ^{2}\hbar^{2}$.

Since the spacetime has a rainbow metric, it is natural that the position of
the brick wall is energy dependent, just like the radius of the event horizon
$r_{h}$. In this sense, in eqn. $\left(  \ref{eq:entropy-rad1}\right)  $ the
$u$ derivative acts on not only the integrand of the integral in the square
bracket, but also the lower limit $r_{h}+r_{\varepsilon}$. Focusing on the
possible most divergent parts near the horizon, we have for the atmosphere
entropy%
\begin{align}
S_{rad}  &  \sim\frac{M}{16\pi^{4}}\int duu^{2}s\left(  u\right)  h^{-\frac
{2}{3}}\left(  \frac{uT_{0}}{m_{p}}\right)  \left[  1-\frac{10T_{0}u}{9m_{p}%
}h^{\prime}\left(  \frac{uT_{0}}{m_{p}}\right)  h^{-1}\left(  \frac{uT_{0}%
}{m_{p}}\right)  \right]  \frac{1}{r_{\varepsilon}}\nonumber\\
&  -\frac{1}{24\pi^{4}}\int dus\left(  u\right)  u^{3}\frac{1}{r_{\varepsilon
}}\frac{dr_{h}}{du}+\frac{M}{48\pi^{4}}\int dus\left(  u\right)
u^{3}h^{-\frac{2}{3}}\left(  \frac{uT_{0}}{m_{p}}\right)  \frac{d}{du}\left(
\frac{1}{r_{\varepsilon}}\right) \nonumber\\
&  -\frac{M}{288\pi^{5}}\int duu^{3}s\left(  u\right)  h^{-\frac{7}{3}}\left(
\frac{uT_{0}}{m_{p}}\right)  h^{\prime}\left(  \frac{uT_{0}}{m_{p}}\right)
\frac{m_{p}}{r_{\varepsilon}^{2}}-\frac{M}{48\pi^{4}}\int dus\left(  u\right)
u^{3}h^{-\frac{2}{3}}\left(  \frac{uT_{0}}{m_{p}}\right)  \frac{dr_{h}}%
{du}\frac{1}{r_{\varepsilon}^{2}}. \label{eq:entropy-rad2}%
\end{align}
It would appear that the most divergent terms are these proportional to
$r_{\varepsilon}^{-2}$. However, it can be shown from eqn. $\left(
\ref{eq:horizon-radius}\right)  $ that the two terms in the last line of eqn.
$\left(  \ref{eq:entropy-rad2}\right)  $ cancel against each other, leaving
only the most divergent terms proportional to $r_{\varepsilon}^{-1}$.

To determine how $r_{\varepsilon}$ depends on $E$, one could introduce the
proper length for $r_{\varepsilon}$ in the rainbow metric $\left(
\ref{eq:FF-Rainbow-BH}\right)  $:%

\begin{equation}
\varepsilon=\int_{r_{h}}^{r_{h}+r_{\varepsilon}}\sqrt{g_{rr}}dr=\frac
{r_{\varepsilon}}{g\left(  E/m_{p}\right)  }.
\end{equation}
Now consider the AC dispersion relation where $f\left(  x\right)  =1$. In this
case, eqn. $\left(  \ref{eq:goverf}\right)  $ gives
\begin{equation}
\varepsilon=r_{\varepsilon}\frac{f\left(  E/m_{p}\right)  }{g\left(
E/m_{p}\right)  }=r_{\varepsilon}h^{-\frac{1}{3}}\left(  \frac{uT_{0}}{m_{p}%
}\right)  . \label{eq:Proper-Distance}%
\end{equation}
One natural assumption is that $\varepsilon$ does not depend on $E$. Under
this assumption, the most divergent part of the atmosphere entropy near the
horizon becomes%
\begin{equation}
S_{rad}\sim\frac{M}{16\pi^{4}\varepsilon}\int_{0}^{u_{\max}}du\frac
{u^{2}s\left(  u\right)  }{h\left(  \frac{uT_{0}}{m_{p}}\right)  }-\frac
{1}{384\pi^{5}}\frac{m_{p}}{\varepsilon}\int_{0}^{u_{\max}}d\tilde{u}%
\frac{u^{3}s\left(  u\right)  h^{\prime}\left(  \frac{uT_{0}}{m_{p}}\right)
}{h^{2}\left(  \frac{uT_{0}}{m_{p}}\right)  }. \label{eq:entropy-rad3}%
\end{equation}
Since $\varepsilon$ is assumed to be independent of $E$, one way to understand
the value of $\varepsilon$ is letting $S_{rad}$ recover the Bekenstein-Hawking
entropy in the usual case, where $h\left(  x\right)  =1$ and $u_{\max}=\infty
$. Thus, we have for $\varepsilon$%
\begin{equation}
\varepsilon=\frac{\hbar}{720\pi M}.
\end{equation}
In this case, for $M\gg m_{p}$ eqn. $\left(  \ref{eq:entropy-rad3}\right)  $
becomes
\begin{equation}
S_{rad}\sim\frac{A}{4\hbar}+\frac{45\left(  3+n\right)  \eta}{128\pi^{5}%
}\left(  \frac{4\pi A}{\hbar}\right)  ^{\frac{2-n}{2}}\int_{0}^{\infty
}dus\left(  u\right)  u^{n+2}, \label{eq:entropy-rad4}%
\end{equation}
where use eqn. $\left(  \ref{eq:hFunction-Small}\right)  $ for $h\left(
x\right)  $. From eqns. $\left(  \ref{eq:Entropy-SC-Small}\right)  $ and
$\left(  \ref{eq:entropy-rad4}\right)  $, it shows that the leading rainbow
corrections to $S_{BH}$ and $S_{rad}$ are both proportional to $A^{\frac
{2-n}{2}}$ in the cases with $n\neq2$. However, the logarithmic divergence
does not appear in $S_{rad}$ for the $n=2$ case, which would imply that
atmosphere entropy could not solely account for the entropy of the black hole.

\section{Discussion and Conclusion}

\label{Sec:Con}

In \cite{RB-SF}, the thermodynamics of a SF rainbow Schwarzschild black hole
was considered. The minimum masses $M_{cr}$ and final temperatures
$T_{BH}^{cr}\ $for the AC dispersion relation with different values of $\eta$
and $n$ were listed in TABLE \ref{tab:two}. Comparing with TABLE
\ref{tab:one}, we find that the behaviors of SF and FF rainbow Schwarzschild
black holes during the final stage of evaporation process are different for
the scenarios with $\eta<0$ and $\frac{2}{3}\leq n\leq2$. Specifically, in the
case with $\eta<0$ and $n=\frac{2}{3}$, a remnant exists for the FF black hole
while it does not for the SF one. In the case with $\eta<0$ and $\frac{2}%
{3}<n<2$, $M_{cr}>0$ and $T_{BH}^{cr}$ is finite for the FF black hole while
$M_{cr}=0$ and $T_{BH}^{cr}=\infty$ for the SF one. In the case with $\eta<0$
and $n=2$, both SF and FF black holes have remnants in their final stages
while $T_{BH}^{cr}$ is finite for the FF one and infinity for the SF one. On
the other hand, TABLEs \ref{tab:one} and \ref{tab:two} show that the behavior
of a FF rainbow black hole appears amazingly similar to that of a SF one,
except for the values of $n$ at which stable remnants occur. For a SF black
hole, the remnant occurs at somewhat higher values of $n$. These similarities
show that the black hole thermodynamics in the rainbow gravity is kind of
independent of the frames used to obtain the rainbow metrics, which hints that
the Gravity's rainbow scenario has some degree of universality.

\begin{table}[h]
\begin{center}
$%
\begin{tabular}
[c]{|c|c|c|}\hline
& $M_{cr}$ & $T_{BH}^{cr}/m_{p}$\\\hline
$\eta=0$ & $0$ & $\infty$\\\hline
$\eta>0$ & $0$ & $\frac{\eta^{-1/n}}{4\pi}$\\\hline
$\eta<0,0<n<2$ & $0$ & $\infty$\\\hline
$\eta<0,n=2$ & $\frac{m_{p}\left\vert \eta\right\vert ^{\frac{1}{2}}}{2}$ &
$\infty$\\\hline
$\eta<0,n>2$ & $\frac{m_{p}}{2\tilde{y}_{0}}$ & $\frac{\tilde{x}_{0}}{4\pi}%
$\\\hline
\end{tabular}
\ $
\end{center}
\caption{The values of $M_{cr}$ and $T_{BH}^{cr}/m_{p}$ for a SF rainbow
Schwarzschild black hole.}%
\label{tab:two}%
\end{table}

In this paper, we considered FF rainbow black holes, and analyzed the effects
of rainbow gravity on the temperature, entropy and atmosphere entropy of a FF
rainbow Schwarzschild black hole. After the metric of a FF rainbow black hole
were proposed, we then used the Hamilton-Jacobi method to compute the
effective Hawking temperature $T_{eff}$ of the rainbow black hole, which
depends on the energy $E$ of emitted particles. By relating the momentum $p$
of particles to the event horizon radius $r_{h}$ of the black hole, the
temperature of a FF rainbow Schwarzschild black hole was obtained. Focusing on
the AC dispersion relation, we computed their minimum masses $M_{cr}$ and
final temperatures $T_{BH}^{cr}\ $for different values of $\eta$ and $n$. All
the results were listed in TABLE \ref{tab:one}. In addition, a non-vanishing
minimum mass indicates the existence of the black hole's remnant, which could
shed light on the \textquotedblleft information paradox\textquotedblright. In
section \ref{Sec:ERSBH}, the atmosphere entropy of a massless scalar field in
a FF rainbow Schwarzschild metric was calculated in the brick wall model.

\bigskip

\bigskip

\begin{acknowledgments}
We are grateful to Houwen Wu and Zheng Sun for useful discussions. This work
is supported in part by NSFC (Grant No. 11005016, 11175039 and 11375121) and
the Fundamental Research Funds for the Central Universities.
\end{acknowledgments}

\end{document}